\documentclass{article}

\usepackage{arxiv}

\newcommand\ADAM{ADAM$_{\textrm{\emph{pro}}}$}

\usepackage{graphics} 
\usepackage{todonotes}
\usepackage{hyperref}
\usepackage{siunitx}
\usepackage{array}
\usepackage{booktabs}
\usepackage{stfloats}
\usepackage{subcaption}
\usepackage{balance}
\usepackage{amssymb}

\title{Towards an All-Purpose Content-Based Multimedia Information Retrieval System}

\author{
  Ralph Gasser \\
  Department of Mathematics and Computer Science\\
  University of Basel\\
  Basel, Switzerland \\
  \texttt{ralph.gasser@unibas.ch} \\
   \And
  Luca Rossetto \\
  Department of Mathematics and Computer Science\\
  University of Basel\\
  Basel, Switzerland \\
  \texttt{luca.rossetto@unibas.ch} \\
  \And
  Heiko Schuldt \\
  Department of Mathematics and Computer Science\\
  University of Basel\\
  Basel, Switzerland \\
  \texttt{heiko.schuldt@unibas.ch} \\
}

\begin{document}

\maketitle

\begin{abstract}
The growth of multimedia collections -- in terms of size, heterogeneity, and variety of media types -- necessitates systems that are able to conjointly deal with several forms of media, especially when it comes to searching for particular objects. However, existing retrieval systems are organized in silos and treat different media types separately. As a consequence, retrieval across media types is either not supported at all or subject to major limitations.

In this paper, we present vitrivr, a content-based multimedia information retrieval stack. As opposed to the keyword search approach implemented by most media management systems, vitrivr makes direct use of the object's content to facilitate different types of similarity search, such as Query-by-Example or Query-by-Sketch, for and, most importantly, across different media types -- namely, images, audio, videos, and 3D models. 
Furthermore, we introduce a new web-based user interface that enables easy-to-use, multimodal retrieval from and browsing in mixed media collections. The effectiveness of vitrivr is shown on the basis of a user study that involves different query and media types.
To the best of our knowledge, the full vitrivr stack is unique in that it is the first multimedia retrieval system that seamlessly integrates support for four different types of media. As such, it paves the way towards an all-purpose, content-based multimedia information retrieval system.
\end{abstract}

\section{Introduction}
\label{sec:introduction}

As media collections grow larger and become more diverse, the quest for accessing the knowledge contained within these collections becomes more arduous. This is mainly due to the lack of proper tools for satisfying a particular information need. The classical approach of annotating media objects and retrieving them later based on this metadata has several shortcomings. Firstly, the sheer amount of data and the pace at which multimedia collections grow makes the laborious task of prior annotation ever more daunting. Secondly, textual descriptions tend to be subjective due to personal experience, expertise, language, and culture. And thirdly, it is difficult to describe temporal evolution, e.g., in videos, in a way that enables others to retrieve the desired object later, i.e., to anticipate all future possible searches at the time an object is annotated. In order to overcome these limitations, retrieval systems need to take the objects' content into account. However, most existing content-based multimedia retrieval systems only address a single media type and do not support the search in several or even across different modalities.

In this paper we present an extended version of vitrivr -- a scalable, open source, content-based multimedia information retrieval stack~\cite{Rossetto2016}. vitrivr is the successor of the IMOTION system~\cite{rossetto2015imotion}, which has originally been designed for multimedia retrieval in large video collections. The work described herein directly builds on these previous efforts. We leverage \ADAM~\cite{Giangreco2016}, a storage engine that facilitates fast and scalable k-Nearest Neighbours (kNN) look-up in high-dimensional vector spaces for the purpose of multimedia retrieval, and Cineast~\cite{Rossetto2014}, a modular feature extraction and retrieval engine developed for video retrieval. As part of our work, we have integrated different, media type specific content-based retrieval techniques into Cineast so as to contrive a solution that is capable of managing and searching not only in video compilations, but large, mixed multimedia collections. 

%\todo[inline]{HS: should we also add a sentence on vitrivr's history, i.e., introduce IMOTION (like in the ICMR demo paper)?}

The contribution of the paper is twofold: First, we introduce the architecture and the supported features of vitrivr, our scalable content-based multimedia information retrieval stack. Second, we show the effectiveness of vitrivr by means of a user study that considers a very heterogeneous set of search tasks.

The remainder of this paper is structured as follows: Section~\ref{sec:related_work} surveys related work. Section~\ref{sec:retrieval_methods} outlines some of the retrieval techniques applied by vitrivr and Section~\ref{sec:implementation} describes implementation aspects. Section~\ref{sec:evaluation} presents the evaluation of the multimodal retrieval support in vitrivr and discusses the results. Section~\ref{sec:conclusion} concludes.

\section{Related Work}

\begin{figure*}[tp]
\begin{subfigure}[b]{.19\textwidth}
  \centering
  \includegraphics[width=.95\textwidth]{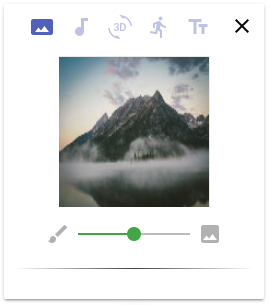}
  \caption{Visual QbE}
  \label{fig:retmet01}
\end{subfigure}
\begin{subfigure}[b]{.19\textwidth}
  \centering
  \includegraphics[width=.95\textwidth]{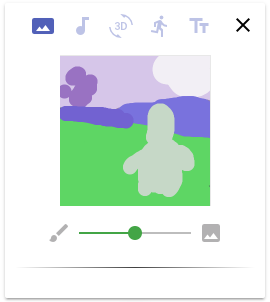}
  \caption{Visual QbS}
  \label{fig:retmet02}
\end{subfigure}
\begin{subfigure}[b]{.19\textwidth}
  \centering
  \includegraphics[width=.95\textwidth]{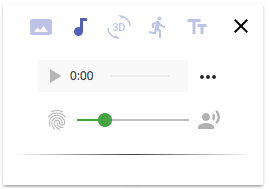}
  \caption{QbE for audio}
  \label{fig:retmet03}
\end{subfigure}
\begin{subfigure}[b]{.19\textwidth}
  \centering
  \includegraphics[width=.95\textwidth]{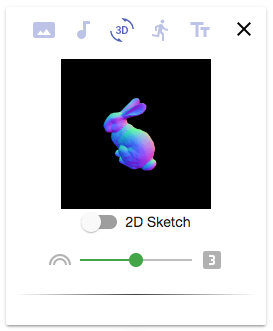}
  \caption{QbE for 3D models}
  \label{fig:retmet04}
\end{subfigure}
\begin{subfigure}[b]{.19\textwidth}
  \centering
  \includegraphics[width=.95\textwidth]{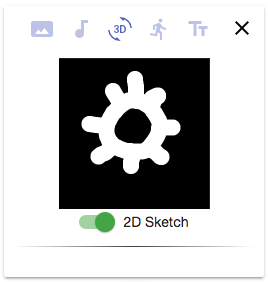}
  \caption{QbS for 3D models}
  \label{fig:retmet05}
\end{subfigure}
\caption{Illustration of how the different retrieval methods present themselves in the user interface. We have query terms for QbE (\ref{fig:retmet01}) and QbS (\ref{fig:retmet02}) of videos and images, QbE of audio (\ref{fig:retmet03}) and QbE (\ref{fig:retmet04}) and QbS (\ref{fig:retmet05}) of 3D models. In every case, the user can either select or create a reference document that is later used for look-up.}
\label{fig:retrievalmethods}
\end{figure*}

\label{sec:related_work}
Currently, there seems to be very little work on integrated solutions for content-based retrieval of different types of media. Most research in the field focuses on a particular modality like audio, video, or 3D or even further specialized subdomains, like for example, music, speech, or environmental sounds for audio. 

Some examples of general-purpose retrieval systems involve QBIC~\cite{Flickner1995} and MUVIS~\cite{Kiranyaz2003}, both of which support retrieval of images and video. Only MUVIS, however, has added support for audio both stand-alone and interlaced with video. Moreover, both systems are not available publicly.

\subsection{Content-Based Image Retrieval}
Early work on image matching and retrieval started in the late 1970's and it has become a fundamental aspect of many problems in computer vision. General-purpose color-based CBIR systems very often employ histograms in different color spaces, color layout and region-based search, or a combination thereof \cite{Mistry2013}. Typical techniques for identifying shapes involve edge histograms or image moments, like for instance centroid distances \cite{Mistry2013,Zhang2002}.
%\todo[inline]{HS: it is a silly question, but it is not fully clear to me: when are terms used in \emph{italics}, when in normal font -- and what is the semantics of terms in italics? For instance: SIFT is not in italics while \emph{BoW} is.}

More recent developments in image retrieval gave rise to advanced techniques like SIFT \cite{Lowe2004}, SURF \cite{Bay2008} for local descriptors and VLAD \cite{Jegou2010} or Fisher Vectors \cite{Perronnin2006} for aggregation. Once local feature descriptors have been obtained by means of SIFT, SURF or a similar approach, it is also possible to apply a Bag of Words (BoW) model to create a global, aggregated feature vector \cite{Yang2007,Liu2013,Ahmad2013}. 

\subsection{Content-Based Music Retrieval}
Content-based Music Retrieval (CBMR) or just Music Information Retrieval (MIR) is a multidisciplinary field that straddles different domains ranging from computer science to psychology. There is a large community surrounding MIR organized in the \emph{International Society for Music Information Retrieval (ISMIR)}\footnote{\url{http://www.ismir.net}}, which holds the annual MIREX evaluation campaign for MIR algorithms~\cite{Downie2010}.

Generally, MIR tasks can be characterized by either their specificity and their granularity. Based on these two dimensions, \cite{Grosche2012} classifies existing Query-by-Example techniques for music into four larger categories: \emph{audio identification (fingerprinting)}, \emph{audio matching}, \emph{version identification} and \emph{category-based retrieval}.

The problem of audio fingerprinting consists in finding exact matches given a short segment of music, that is, the recording the segment belongs to. This problem has been largely solved and different methods have been developed and are being used in commercial applications. Prominent examples include Shazaam\footnote{\url{https://www.shazam.com}} or \emph{Mel-frequency Cepstrum Coefficients (MFCC)}~\cite{Foote1997,Wang2006,Sigurdson2006}. In contrast, even though there is a lot of ongoing research, no best practices have evolved yet for audio matching or version identification and a lot of working examples trade retrieval accuracy for scalability or vice versa. These two tasks consist in finding different variations of a given piece of music, for example, a live recording, a cover version, or a remix. The notion of similarity, hence, becomes more fuzzy in these cases. Generally, chroma-based features, like \emph{Pitch Class Profiles (PCP)}~\cite{Fujishima1999,Gomez2006a} or variations thereof~\cite{Kurth2008} were shown to be well suited for these MIR tasks, but some systems also exploit rhythm or melody~\cite{Salamon2012}. For instance, \cite{Grosche2012a} proposes a version identification scheme based on previous work by Casey et al.~\cite{Casey2006,Casey2008}. The proposed technique involves comparing pitch class profiles for overlapping shingles (audio fragments) of fixed length. The authors were able to demonstrate that this technique scales well, especially when combined with \emph{Locality Sensitive Hashing (LSH)}~\cite{Indyk1998}.

\subsection{Content-Based 3D Model Retrieval}
Similarity of 3D models can be assessed by means of many different descriptors of which feature vectors, histograms, and statistical moments are only three examples. Multiple surveys \cite{Tangelder2007,Bustos2006} list and classify the many methods for content-based 3D model retrieval. 

In \cite{Saupe2001,Kazhdan2003}, a linear combination of \emph{spherical harmonics} is used to obtain a function that approximates the 3D model's surface. The weight coefficients of this function serve as components in a feature vector. On the other hand, \cite{Chen2003} proposes \emph{light field descriptors} for 3D models, which are based on projections of the 3D model onto the faces of a circumscribing dodecahedron. Subsequently, classical CBIR techniques can be applied on the resulting images to extract feature descriptors, namely calculation of \emph{Zernike moments} and \emph{Fourier descriptors} for the resulting shape. 

\section{Retrieval Methods}
\label{sec:retrieval_methods}
The vitrivr system allows for \emph{Query-by-Example (QbE)} and \emph{Query-by-Sketch (QbS)}. The latter is only supported for visual modalities, that is, videos, images, and 3D models. The QbE paradigm takes a reference document, for instance an example image or a short audio snippet, and tries to find documents in the corpus that are similar to the reference. In contrast to this is the \emph{More-Like-This} query mode bases queries on previously retrieved documents which are already known to the system. QbS is a special case of QbE in which the reference document is a hand-drawn sketch produced by the user. For this case, the user interface provides a simple canvas that allows the user to directly create these sketches. For videos, there is an additional variant of QbS which enables a user to sketch motion paths, which we refer to as \emph{Query-by-Motion} (QbM).

Figure \ref{fig:retrievalmethods} illustrates how  users can interact with the different retrieval methods in the user interface. Each Figure \ref{fig:retmet01} to \ref{fig:retmet05} depicts one query term, each using a different type of either selected or hand-crafted reference document.

\subsection{Retrieval of Images and Videos}
The retrieval of visual modalities in images and video is largely based on the pre-existing capabilities of Cineast. Most of the features employed by the original version can be directly applied to still images as the state-of-the-art in video retrieval can be, with a few exceptions, reduced to the use of image retrieval techniques in combination with keyframing. All the original Cineast features are also supported by the current version.

In the extended system, we have added feature modules based on SURF~\cite{Bay2008} and HOG~\cite{McConnell1986}, combined with a simple BoW model to further the support for exact matches in both images and videos. The codebooks for these tasks were derived from the MIR Flickr 25k \cite{Huiskes2008} collection.

\subsection{Retrieval of Music}
The new feature modules for music retrieval are largely based on HPCP \cite{Fujishima1999,Gomez2006a} and CENS~\cite{Mueller2005} features combined with the shingling approach proposed by~\cite{Grosche2012a}. These features can be used for audio matching tasks. Moreover, we have added some fingerprinting methods, namely a feature based on MFCC~\cite{Foote1997} and one inspired by the Shazam algorithm~\cite{Wang2006}. 

As one can see in Figure \ref{fig:retmet03}, the user interface allows a user to upload short audio segments of a few seconds to be used as reference documents. Furthermore, the UI can be used to specify the type of query that should be executed along the dimensions mentioned in Section \ref{sec:related_work}, namely audio fingerprinting, version identification, or audio matching. These settings influence the feature modules that are being executed by Cineast. For pure fingerprinting tasks, only the fingerprinting modules will be used whereas for audio matching, the CENS and HPCP modules are used and fingerprinting modules are left out.

The current version of Cineast only includes audio features that can be used for music retrieval. However, it is possible to add more feature modules to add support for speech or general purpose audio.

\subsection{Retrieval of 3D Models}
The new feature modules for QbE in 3D model retrieval are based on spherical harmonics as proposed by \cite{Saupe2001,Kazhdan2003}. For QbS, the users can draw projections of the object as perceived when looking at the model from a specific angle. For similarity search, we compare Fourier and Zernike coefficients calculated for the model's projections onto a circumscribing dodecahedron. This is referred to as light field descriptor and was proposed by \cite{Chen2003}.

The user interface allows the user to either upload a reference model for QbE or to sketch a 2D silhouette of the desired object and retrieve 3D models based on its shape. See Figures \ref{fig:retmet04} and \ref{fig:retmet05} for how this looks like.

\section{Implementation}
\label{sec:implementation}
The entire vitrivr stack consists of three components: a web-based user interface called Vitrivr-NG, the feature extraction and query processing engine Cineast and \ADAM, a storage layer for high-dimensional feature vectors. In addition, a web server is required in order to host and serve the multimedia files and derivatives like thumbnails or clips. 

%\todo[inline]{HS: the term `system context' in the image caption is not fully clear to me. Also, the term vitrivr does not appear on the figure. Could vitrivr be added and `system context' be explained (or is it rather system architecture instead of system context)?}

\begin{figure*}[tb]
  \centering
  \includegraphics[width=0.9\textwidth]{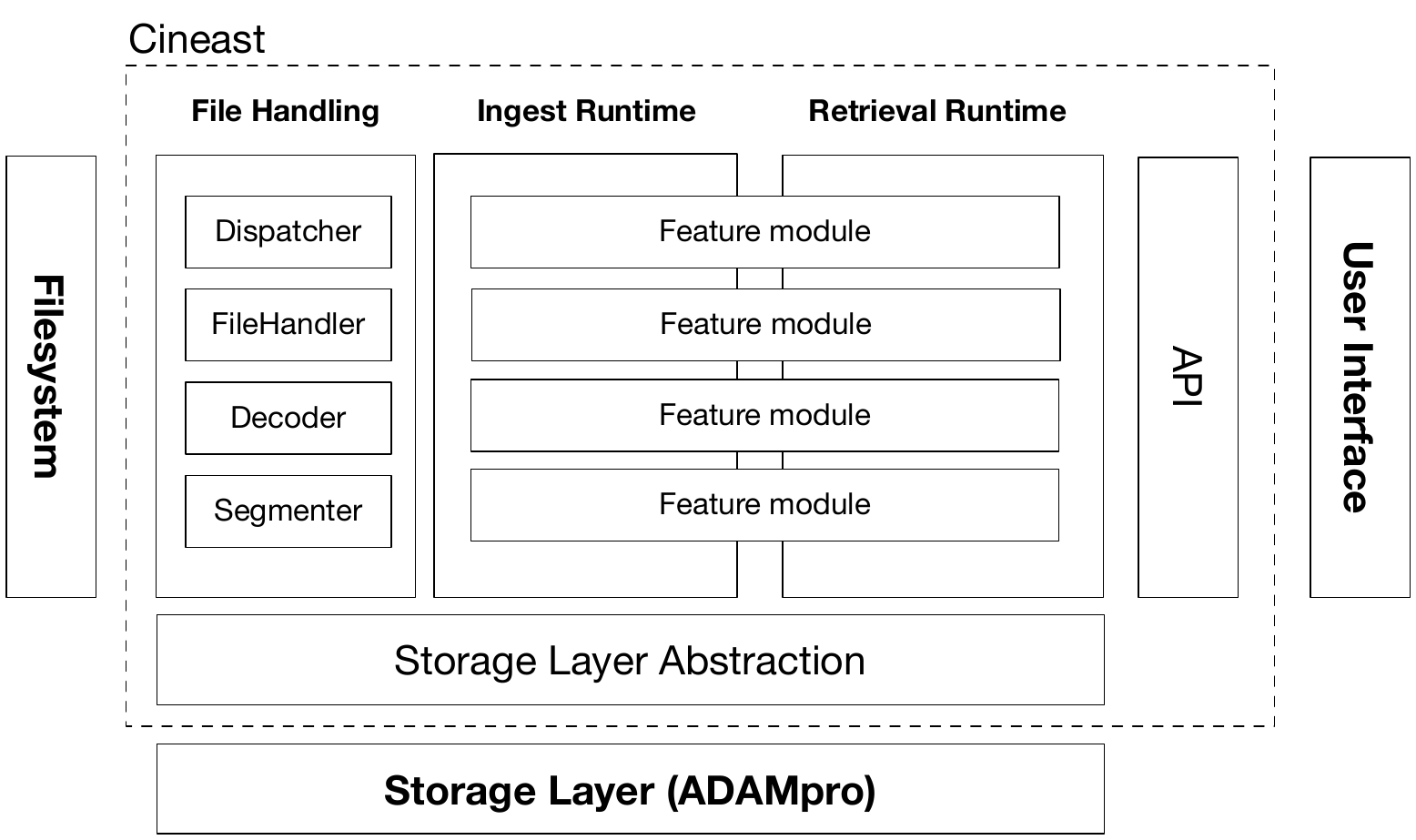}
  \caption{Illustration of Cineast's system architecture. The main modules are the file handling module, the ingest runtime, and the retrieval runtime. These three modules facilitate the offline ingest and the online retrieval workflow supported by the system. Outside the system context lies the storage layer for feature vectors and the user interface.}
   \label{fig:cineastarchitecture}
\end{figure*}

\subsection{The Retrieval Engine}
Cineast is a modular feature extraction and multimedia retrieval engine implemented in Java and forms the core of the entire stack. Its architecture is illustrated in Figure~\ref{fig:cineastarchitecture}. Cineast supports two types of workflows: The offline \emph{ingest workflow} and the online \emph{retrieval workflow}. The offline workflow consists in decoding and segmentation of multimedia files and Cineast  includes support for a wide variety of formats through libraries like FFMPEG\footnote{\url{https://www.ffmpeg.org}} and TwelveMonkeys\footnote{\url{https://github.com/haraldk/TwelveMonkeys}}. In the process, the derived segments are ultimately handed to an extraction pipeline where they are processed by different \emph{feature modules}. The online retrieval workflow parses queries submitted by users and utilizes the provided \emph{reference documents} (e.g., a short audio segment or a sketch) to derive feature descriptors also using the aforementioned feature modules. The feature modules thus provide the main functionality of Cineast both during the online and the offline workflow. They derive feature descriptors from a segment, generate feature vectors and then use the storage layer to either persist them (offline) or perform a look-up (online).

The query functionality of Cineast can be accessed through a WebSocket and a RESTful API. The API supports different types of actions, like simple ID or keyword-based look-ups, kNN search based on a provided reference document or kNN search using an existing entry in the database (``More-Like-This'' queries).

Cineast's modular architecture allows for easy extension in terms of supported features just by adding new feature modules and composing these modules into different feature categories. As part of the work reported in this paper, we have added 14 new feature modules in order to support the query modes described in Section~\ref{sec:retrieval_methods}.

\subsection{The Storage Layer}
\ADAM\ is a database system that is able to persistently store and retrieve multimedia objects on a large scale. Most importantly, it allows for efficient kNN search in high-dimensional vector spaces, which is crucial for content-based multimedia retrieval. It employs different exact and approximative indexing strategies like \emph{Spectral Hashing (SH)}~\cite{Weiss2008}, LSH~\cite{Indyk1998} and \emph{Vector-Approximation (VA) files}~\cite{Weber1998}. In addition, \ADAM\ also supports the storage of ordinary entities with textual, numerical, and temporal information. We refer to~\cite{Giangreco2016} for more information on \ADAM.

\ADAM\ acts as the storage layer in the vitrivr stack. It contains all the information on multimedia objects, segments thereof, and the extracted feature vectors. This also includes technical information about files and file metadata. Queries formulated by the Cineast engine are delegated to \ADAM, which executes them and returns the results.

\subsection{The User Interface}
\begin{figure*}[t]
  \centering
  \includegraphics[width=1.0\textwidth]{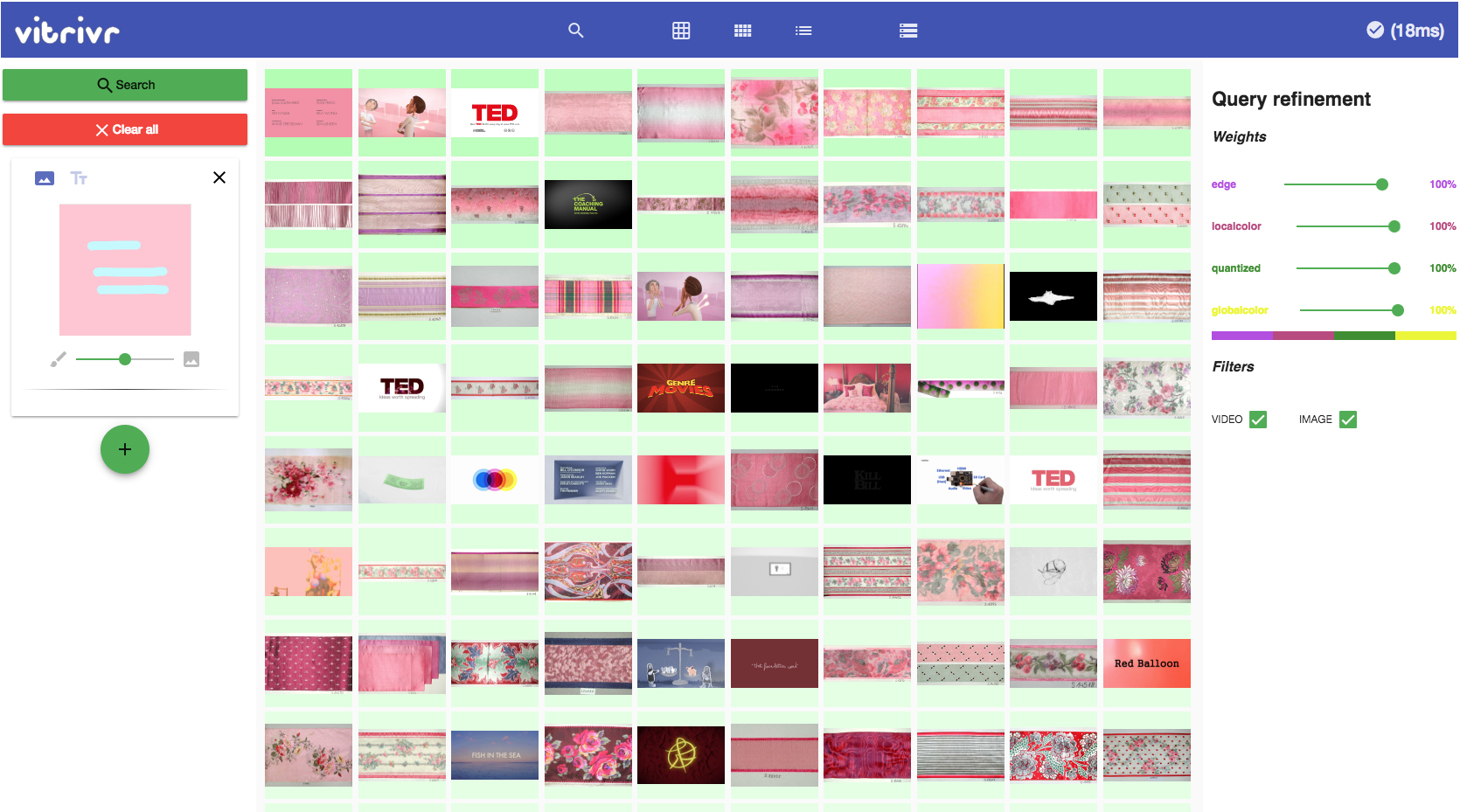}
  \caption{Impression of the web-based user interface for the system. On the left-hand part, users can formulate queries. Results are presented in the middle part of the UI. The right-hand part of the UI can be used to refine the result set after a query has been issued, by filtering media types or weighting features differently. The green color coding is used to indicate relevance.}
   \label{fig:ui}
\end{figure*}

The web-based user interface of vitrivr integrates query and display mechanisms for the different media types. It was built using Angular \footnote{\url{https://angular.io}} and has been written in TypeScript 2.1. Communication with Cineast takes place through the aforementioned RESTful and WebSocket API's. An impression of the UI is given in Figure~\ref{fig:ui}.

The main design goal for the UI was to build a modular, extensible, web-based user interface that maintains the functionality of the original version~\cite{Rossetto2016} and extends it, so as to support queries for and across different modalities. This includes not only building such queries but also presenting the different types of results in a consistent manner. The Angular framework is well suited to that end, because its modular architecture allows us to easily add and remove components as the stack evolves.

Mainly, the user interface assists the user in composing queries from different building blocks. The first block is called \emph{query component}. Each query component consists of multiple \emph{query terms} that can be toggled. A query component must at least contain one active query term and only one instance of a query term of the same type can be active per component. The individual query terms differ in the kind of reference document that is being used. For example, the image query term enables users to upload a reference image or sketch one themselves whereas the audio query term allows the user to upload short audio clips.  Some examples of query terms supported by vitrivr are depicted in Figure~\ref{fig:retrievalmethods}. Upon execution, the query terms within a query component are connected by a logical AND relationship whereas different query components are connected by a logical OR. This simple model allows the user to formulate complex queries within and across different modalities. The scheme can also be easily extended with new types of queries, like for instance, motion sketches or search for textual data, by just adding a new type of query term.

\section{Evaluation}
\label{sec:evaluation}
We evaluate vitrivr's retrieval effectiveness in terms of utility for the end user. The evaluation is based on two test sets (A and B) that comprise twelve similar scenarios each. There are three scenarios per domain, that is, image (1 to 3), audio (4 to 6), video (7 to 9) and 3D model retrieval (10 to 12) and each scenario focuses on different aspects within the respective domain. A scenario comprises a simple objective (information need), which the users are expected to carry out using the web-based user interface. Each scenario is described textually and in some cases the textual description is supported by an illustrative image, for example depicting the scene the user should find in a video. Unless otherwise stated, the users are not allowed to use these helper images directly as reference for the query. However, they are allowed to, for example, use a helper image as template for creating their own sketch. Figure~\ref{fig:evaluation} gives some examples of these helper images. Some scenarios also comprise additional material the participants can use to perform the task at hand. For instance, we provide short audio snippets for the audio retrieval tasks.

In general, users are free to execute as many queries as they please until they are either satisfied with the result or they decide to give up. Here, we apply the principle that searching is an iterative process~\cite{Jansen2010}. Unless otherwise stated, users are allowed to leverage all of the UI's capabilities, namely QbS, QbE (query based on external document), and More-Like-This (query based on previously retrieved document) queries. Additionally, they may also use the refinement functionality provided by the UI. Applying the \emph{principle of the least effort}~\cite{Jansen2010,Zipf1994}, we expect the users to take the course of action that they believe to be connected with the least expenditure. Once a user has obtained and accepted a result for a scenario, they are required to rate the top $15$ documents on the following four point scale:

\begin{itemize}
\item[0:] Resulting document is considered  irrelevant.
\item[1:] Resulting document is considered slightly relevant.
\item[2:] Resulting document is considered very relevant.
\item[3:] Resulting document is considered highly relevant, close to identity.
\end{itemize}

The relevance judgments are then aggregated into MAP, MRR, NDCG@15 and p@15 values per scenario. For binary metrics, ratings of $2$ and $3$ are considered to be hits and values of $0$ and $1$ are considered to be misses. Furthermore, we decide for each scenario whether the user was able to fulfill the objective (success rate) based on the presence of at least one high-relevance rating and how many queries were required on average.

\subsection{Test Collections}
For the evaluation, we have assembled our own test collections. We selected random items from the Freemusicarchive, Pixabay, and Thingiverse. In order to construct examples for audio matching, we included specific instances of contemporary and popular music from various sources.
Table~\ref{tab:collections} lists all the collections that were used in the evaluation.

\subsection{Results}
The outcome of the user-driven evaluation is summarized in Table~\ref{tab:ude-results}. In total, $25$ datasets were gathered. 13 participants did test set A and the other twelve worked through test set B.

\begin{table*}[bp]
  \caption{List of all the collections used during the evaluation. The table contains information regarding the size and source.}
  \label{tab:collections}
  \begin{tabular*}{\textwidth}{c @{\extracolsep{\fill}} llll}
    \hline
    \textbf{Domain} & \textbf{Name} & \textbf{Entries} & \textbf{References} \\
    \hline
    Image & Pixabay & 164512 & https://pixabay.com \\   \hline
    Video & OSVC & 200 & See \cite{Rossetto2015} \\  \hline
     Audio & Freemusicarchive & 4335 & https://freemusicarchive.org \\   \hline
    Audio & Misc. sources & 62 & - \\  \hline
    3D & NTU 3D (1-4) & 4003 & See \cite{Chen2003b} \\  \hline
    3D & Thingiverse & 8966 & https://www.thingiverse.com \\  \hline
  \end{tabular*}
\end{table*}

\begin{table*}[bp]
  \caption{Averaged, per scenario results of the user-driven evaluation. The success rate indicates in how many cases at least one highly relevant item was obtained. The last column indicates how many queries were executed on average.}
  \label{tab:ude-results}
  \setlength{\tabcolsep}{0.95em}
  \centering
  \begin{tabular}{ |l|c|c|c|c|c|c|c|c|c|c|c|c| }
    \toprule
    \textbf{\#} & \multicolumn{2}{c}{NDCG@15} & \multicolumn{2}{c}{p@15} & \multicolumn{2}{c}{MRR} & \multicolumn{2}{c}{MAP} & \multicolumn{2}{c}{Success rate} & \multicolumn{2}{c|}{\# queries}\\
	\midrule
     & \textbf{A} & \textbf{B} & \textbf{A} & \textbf{B} & \textbf{A} & \textbf{B} & \textbf{A} & \textbf{B} & \textbf{A} & \textbf{B} & \textbf{A} & \textbf{B} \\ \hline
      1 & 0.91 & 0.49 & 0.60 & 0.17 & 0.77 & 0.33 & 0.33 & 0.12 & 0.92 & 0.67 & 1.8 & 1.7 \\ \hline
      2 & 0.96 & 0.94 & 0.13 & 0.16 & 0.92 & 1.0 & 0.10 & 0.11 & 0.92 & 1.0 & 2.2 & 1.1 \\ \hline
      3 & 0.67 & 0.87 & 0.11 & 0.08 & 0.63 & 0.80 & 0.08 & 0.08 & 0.69 & 1.0 & 3.3 & 2.2 \\ \hline
      4 & 0.97 & 0.98 & 0.08 & 0.07 & 1.0 & 1.0 & 0.07 & 0.07 & 1.0 & 1.0 & 1.0 & 1.0\\ \hline
      5 & 0.61 & 0.97 & 0.07 & 0.08 & 0.47 & 1.0 & 0.06 & 0.08 & 0.85 & 1.0 & 1.8  & 1.0 \\ \hline
      6 & 0.87 & 0.86 & 0.14 & 0.13 & 1.0 & 0.92 & 0.10 & 0.10 & 1.0 & 1.0 & 1.2 & 1.3 \\ \hline
      7 & 0.85 & 0.30 & 0.11 & 0.03 & 0.56 & 0.17 & 0.08 & 0.03 & 0.77 & 0.33 & 1.7 & 2.3 \\ \hline
      8 & 0.74 & 0.77 & 0.14 & 0.09 & 0.63 & 0.66 & 0.10 & 0.08 & 1.0 & 1.0 & 1.0 & 1.2\\ \hline
      9 & 0.73 & 0.54 & 0.07 & 0.05 & 0.64 & 0.39 & 0.07 & 0.05& 0.92 & 0.67 & 1.7 & 2.0 \\ \hline
      10 & 0.83 & 0.93 & 0.12 & 0.10 & 0.70 & 1.0 & 0.09 & 0.08 & 0.69 & 1.0 & 2.8 & 4.5 \\ \hline
      11 & 0.97 & 0.92 & 0.69 & 0.23 & 1.00 & 1.0 & 0.38 & 0.15 & 1.0 & 1.0 & 2.6 & 1.6 \\ \hline
      12 & 0.89 & 0.90 & 0.26 & 0.34 & 0.77 & 1.0 & 0.16 & 0.20 & 0.77 & 1.0 & 2.2 & 2.5 \\ \hline
  \end{tabular}
\end{table*}

\subsection{Discussion}

In this section, we discuss the evaluation results scenario by scenario. Scenarios 1 and 2 aimed at QbE for images. In the first task, users were free to pick a reference image of their choice whereas for the second task, users were provided with a slightly altered version (blurring, discoloration) of an image that was contained in the collection. Interestingly, not all users considered the results returned in A1 and B1 to be relevant. A success rate of \SI{92}{\%} and \SI{66}{\%}, respectively, is not great for a simple task like this and neither is a MRR of $0.77$ and $0.33$. As it turns out, the results are very dependent on the reference image. This is also reflected by the p@15 value, which is considerably higher for A1 than for B1. Our findings suggest that if the general color setting of the reference image remotely matches an image in the database, the latter is set to rank high in the list of results regardless of whether it depicts the same thing conceptually. Hence, the color features outweigh features that take local structures into consideration.

For scenarios A2 and B2, \SI{92}{\%} and \SI{100}{\%} of the users were able to obtain the copy of the reference image and the MRR of $0.92$ and $1.0$ indicates that the desired image was at rank 1 most of the time. Thus, the system seems to be pretty robust against minor alterations of the reference image. From the p@15 value, we must deduce that a majority of the remaining results were considered to be irrelevant or only slightly relevant. The NDCG@15 implies, however, that the ranking coincided pretty well with the rating of the users.

\begin{figure*}[t]
\begin{subfigure}[b]{.3\textwidth}
  \centering
  \includegraphics[width=.9\textwidth]{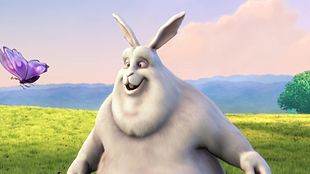}
  \caption{A7, illustrative image}
  \label{fig:eval01}
\end{subfigure}
\begin{subfigure}[b]{.3\textwidth}
  \centering
  \includegraphics[width=.6\textwidth]{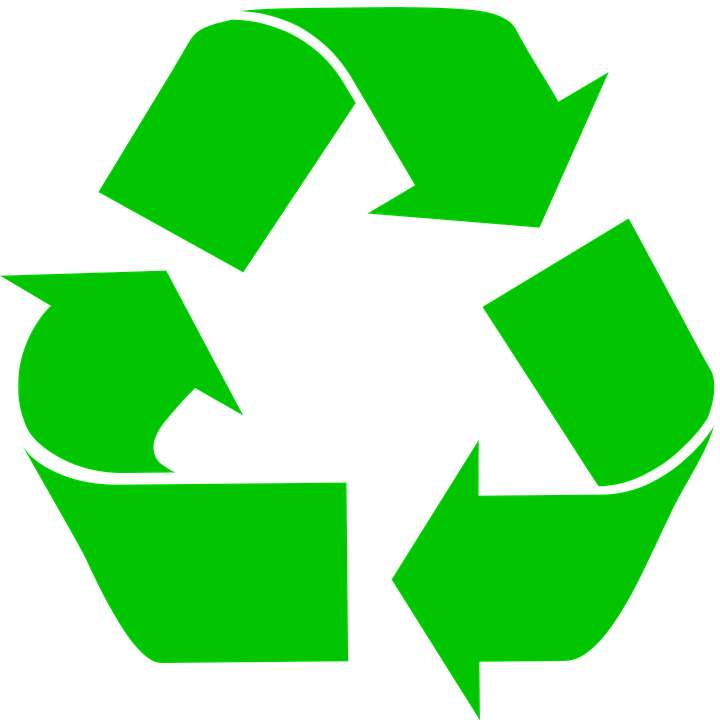}
  \caption{B3, illustrative image}
  \label{fig:eval02}
\end{subfigure}
\begin{subfigure}[b]{.3\textwidth}
  \centering
  \includegraphics[width=.7\textwidth]{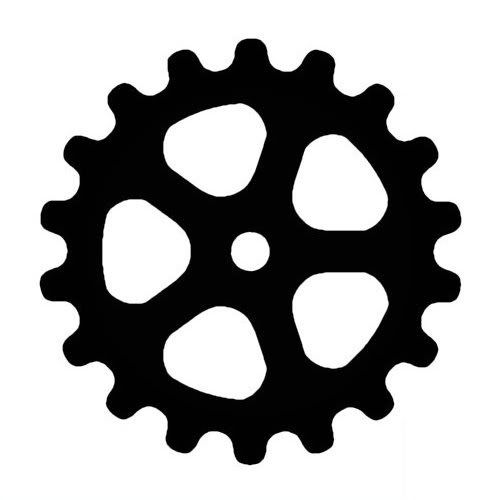}
  \caption{B12, illustrative image}
  \label{fig:eval03}
\end{subfigure}
\begin{subfigure}[b]{.33\textwidth}
  \centering
  \includegraphics[width=.8\textwidth]{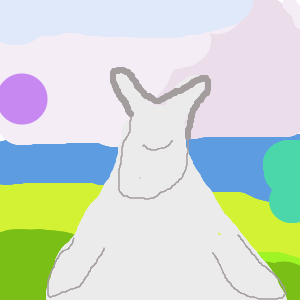}
  \caption{A7, user sketch (\checkmark)}
  \label{fig:eval04}
\end{subfigure}
\begin{subfigure}[b]{.33\textwidth}
  \centering
  \includegraphics[width=.9\textwidth]{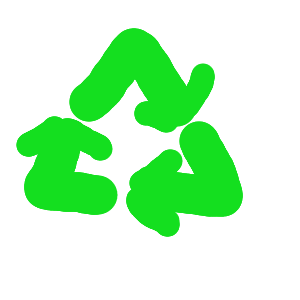}
  \caption{B3, user sketch (\checkmark)}
  \label{fig:eval05}
\end{subfigure}
\begin{subfigure}[b]{.33\textwidth}
  \centering
  \includegraphics[width=.7\textwidth]{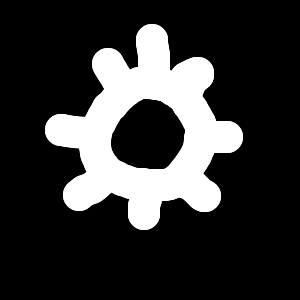}
  \caption{B12, user sketch (\checkmark)}
  \label{fig:eval06}
\end{subfigure}
\caption{A selection of illustrations shown to and sketches produced by the participants of the evaluation. The examples here include QbS tasks for video (A7), still images (B3) and 3D models (B12). All the depicted sketches were conductive to successfully retrieving an object of interest from the corpus.}
\label{fig:evaluation}
\end{figure*}

Scenarios A3 and B3 involved QbS tasks for images. In both cases, the users had to find a particular logo or icon based on a sketch. Figure \ref{fig:evaluation} depicts the helper image (\ref{fig:eval02}) and an example sketch produced by a participant (\ref{fig:eval05}) of scenario B3. Success rates of \SI{69}{\%} and \SI{100}{\%}, respectively, and a MRR of $0.63$ and $0.80$ indicate that the majority of users were able to retrieve the item of interest and that the relevant item was placed in the top half of the result set. However, not all users managed to retrieve it in the case of A3. It is also worth noting that it took the users $3.3$ and $2.2$ queries on average to obtain the results. This is approximately one query more than for scenarios 1 and 2, which is likely due to the fact that most users required multiple attempts at sketching the item of interest. Also, according to feedback, a lot of users employed the More-Like-This functionality to push the desired item from higher ranks to the top.

Scenarios 4 and 5 were pure audio fingerprinting tasks of which A4 and B4 encompassed a simple, \SI{3}{second} music segment as reference document and A5 and B5 encompassed a music excerpt with a mix of white and rose noise. With the exception of A5, the success rate was \SI{100}{\%} for these tasks, which means that the audio segment in question could always be retrieved. The MRR value for these tasks (again with the exception of A5) was $1.0$, indicating that the relevant document had top rank. The remaining results in the top 15 ranks can be considered irrelevant hits, which explains the low p@15 values. As most users seem to have agreed with the proposed ranking (3 for the first item, 0 for the rest) the NDCG@15 tends to be close to $1.0$. In summary, one can state that the fingerprinting works and exhibits some robustness to noise.

Scenarios A6 and B6 were audio matching scenarios. In both scenarios, the users were supposed to find the original version and a cover version of the same musical piece. Again, the high success rate of \SI{100}{\%} for both tasks indicates that at least one of the versions could be retrieved. The MRR here lies between $1.0$ and $0.92$, which indicates a rank between $1$ and $3$ for the first, highly relevant item in the list. The cover version was also retrieved in most cases. Furthermore, some of the other top 15 items were considered to be highly or at least very relevant. Both these facts contribute to a p@15 value between $0.13$ and $0.14$. The NDCG@15 value indicates that the ranking coincides with the user rating in many instances. It is, however, not perfect.

Scenarios A7 and B7 were pure QbS tasks for video in which users were asked to sketch a scene based on a presented helper image. Figure \ref{fig:evaluation} depicts the helper image (\ref{fig:eval01}) and an example sketch produced by a participant (\ref{fig:eval04}) of scenario A7. Judging from the the comparatively low success rates of \SI{73}{\%} and \SI{33}{\%}, respectively, and the low MRR value, this tasks was very challenging for users to complete.
Based on user feedback, especially the B7 reference image used a very disadvantageous color palette which was very difficult to reproduce without the help of advanced painting tools. These examples -- together with A3 and B3 -- confirm the difficulties of QbS, especially for complex imagery. Again, it is worth noting that both A7 and B7 required $1.69$ to $2.25$ queries per image --- more than the QbE-based tasks.

In scenarios A8 and B8, participants were tasked to combine an audio excerpt with a reference image of choice to find a particular scene in a video. The success rate of \SI{100}{\%} is positively surprising, especially as it is higher than for A9 and B9, where users were only allowed to use a reference image alone. This indicates that adding another modality to the mix indeed brings some advantages even though the relative contribution of the audio features depend on the provided reference image. However, the ranking of the results does not always seem to agree with the user ratings as we can read from the NDCG value between $0.74$ and $0.77$ and precision tends to be rather low. The latter can be attributed to the fact that we were actually looking for a particular scene which is unique in the entire collection, both in terms of the visual as well as the auditory part. As in scenarios A4, A5, B4, and B5, the audio fingerprinting feature, which was used most of the time, reliably produces one accurate hit and a lot of seemingly unrelated results.

In scenarios A9 and B9 users had to retrieve a specific scene based on a provided but distorted image (A9) or example image of their choice (B9). Unsurprisingly, the success rate for A9 was almost \SI{30}{\%} higher than for B9. Also, the MRR in both cases was relatively low as it ranged between $0.39$ and $0.64$. This indicates that it was difficult for the users to bring the desired video to the top rank.

Scenarios A10 and B10 tasked the users to find a type of 3D model based on a 2D sketch. For example, in scenario A10, users were asked to retrieve a model of the starship enterprise. From the success rate of \SI{69}{\%} and \SI{100}{\%}, respectively, and the MRR of $0.69$ and $1.0$, we can deduce that most of the users succeeded in finding a relevant model and that if they found it, it was ranked at top position. However, as for all the QbS tasks so far, the number of queries is comparatively higher than for the other tasks. In fact, users required an average of $2.8$ and $4.5$ queries in order to fulfill A10 and B10, respectively.

Scenarios A11 and B11 were QbE tasks for 3D model retrieval involving a provided reference document each. This task was apparently straightforward in both cases. Both scenarios show a \SI{100}{\%} success rate and a MRR value of $1.0$, which means that highly relevant items could always be obtained and were always placed at the top rank. What is interesting, though, is the large discrepancy of p@15 values, which was $0.69$ for A11 and $0.23$ for B11. This can be attributed to the difference of retrieval performance of the spherical harmonics features for different classes of objects as reported by \cite{Tangelder2007,Bustos2006}.

In scenarios A12 and B12, we asked the user to find a described 3D model of interest by whatever means they prefer. They were allowed to use external resources like Google. Interestingly, most users chose the QbS mode here and were able to obtain relevant items in most of the cases. Figure \ref{fig:evaluation} depicts the helper image (\ref{fig:eval03}) and an example sketch produced by a participant (\ref{fig:eval06}) of scenario B12. The scenario stated that the user was supposed to find a 3D model of a gear. The p@15 values of $0.26$ and $0.34$ indicate that additional items were also found that were considered to be marginally relevant. Judging from the NDCG@15, the ranking by vitrivr coincided pretty well with the rating provided by the users.

\section{Conclusion and Future Work}
\label{sec:conclusion}
In this paper, we were able to demonstrate that vitrivr, a software stack originally designed for content-based video retrieval, can be seamlessly extended to support additional modalities like audio, images, or 3D models.

We have added $14$ additional feature modules that describe different properties of images, music, and 3D models and build on ideas from various authors. With this, we  could show that combining modalities, as in the case of video, can have a positive effect on retrieval performance. 

To the best of our knowledge, we have thereby created the first  integrated content-based multimedia retrieval stack --- taking the idea behind \cite{Kiranyaz2003,Perica2000} one step further. Meanwhile, we laid the foundation for future work in the multimedia retrieval domain, as vitrivr can be considered a framework to design, implement, and test new retrieval techniques and adapt them for specific use cases and requirements. The entire vitrivr stack has been made available as open source software and it can be downloaded from GitHub\footnote{\url{https://github.com/vitrivr/}}.

In our future work, we plan to extend the audio retrieval capabilities of vitrivr. The majority of music features that have been added as part of this work are based on chroma and melody. Hence, we will invest in additional features based on, for instance, rhythm and tempo, and in assessing their influence on the retrieval effectiveness for mid-specificity tasks like version identification or audio matching. In addition, we plan to lower specificity beyond that of audio matching and add additional query modes like \emph{Query-by-Humming}. Also, we want to add support for other types of audio like speech or environmental sound.

On the 3D model retrieval side it would be interesting to take the QbS paradigm for 3D models one step further and add support for QbE with arbitrary images. This would allow for completely novel use cases, like finding 3D models that were used in a scene of a rendered video. However, it would also require more advanced image segmentation. Last but not least, we plan on combining the current state of the vitrivr stack with recent development in deep learning. To that end, we have already added an integration layer for TensorFlow\footnote{\url{https://www.tensorflow.org}} based models.

\section*{Acknowledgements}
This  work  was  partly  supported  by the  Swiss  National Science Foundation, project IMOTION (20CH21\_151571).

%\todo[inline]{HS: I have fixed some of the references in the ICMR demo paper. Maybe these references could be moved to this paper as well? We can discuss this tomorrow ...}

\balance{}

\bibliographystyle{unsrt}
\bibliography{bibliography}

\end{document}